\documentclass{epl}
\usepackage{graphicx} \usepackage{subfigure}

\title{Response of a Hexagonal Granular Packing under a Localized
External Force}
\shorttitle{Response of a Hexagonal Granular Packing}

\author{Srdjan Ostojic and Debabrata Panja} \institute{
\inst{1}Institute for
Theoretical Physics, Universiteit van Amsterdam, Valckenierstraat 65,
1018 XE Amsterdam, The Netherlands}


\begin{document}

\maketitle

\begin{abstract} 
We study the response of a two-dimensional hexagonal packing of rigid,
frictionless spherical grains due to a vertically downward point force
on a  single grain at the  top layer.  We use  a statistical approach,
where each configuration  of the contact forces is  equally likely. We
show that  this problem  is equivalent to  a correlated  $q$-model. We
find that the  response displays two peaks which  lie precisely on the
downward lattice directions emanating from the point of application of
the  force.   With  increasing  depth,  the  magnitude  of  the  peaks
decreases, and a central peak develops.  On the bottom of the pile,
only  the middle  peak persists.   The response  of  different system
sizes exhibits self-similarity.
\end{abstract}

Force transmissions in (static) granular packings have attracted a lot
of attention in recent years
\cite{review,Brev,fluct,vanel,qmodel,ned,gold,reyd,serero,atman,exp1a,exp1b,exp2}. Granular
packings are assemblies of macroscopic particles that interact only
via mechanical repulsion effected through physical
contacts. Experimental and numerical studies of these systems have
identified two main characteristics. First, large fluctuations are
found to occur in the magnitudes of inter-grain forces, implying that
the probability distribution of the force magnitudes is rather broad
\cite{fluct}. Secondly, the average propagation of forces --- studied
via the response to a single external force --- is strongly dependent
on the underlying contact geometry \cite{vanel, exp1a, exp1b, exp2}.

The available  theoretical models capture  either one or the  other of
these  two  aspects.  The  scalar $q$-model  \cite{qmodel}  reproduces
reasonably well the observed  force distribution, but yields diffusive
propagation  of  forces,  in  conflict with  experiments  \cite{exp1a,
exp1b}.  Continuum  elastic  and  elasto-plastic  theories  \cite{ned}
predict   responses   in   qualitative  agreement   with   experiments
\cite{gold, reyd, serero, atman}, but they provide a description only at the average
macroscopic  level.  More  ad-hoc ``stress-only''  models  \cite{Brev}
include   structural   randomness,  but   its   consequences  on   the
distribution of forces  are unclear. In other words,  an approach that
produces  both realistic  fluctuations  and propagation  of forces  in
granular materials from the {\it same set of fundamental principles\/}
is still called for.
 
A simple conjecture, which  could provide such a fundamental principle
for all problems of granular  statics, has been put forward by Edwards
years  ago  \cite{edwards,edwards2}.  The  idea  is  to  consider  all
``jammed'' configurations  equally probable. {\it A  priori}, there is
no justification  for such an ergodic hypothesis,  but its application
to  models  of  jamming  and  compaction has  been  rather  successful
\cite{jamming}. Its extension to the forces in granular packings is in
principle   straightforward:   sets  of   forces   belonging  to   all
mechanically stable configurations have equal probability. However, in
an ensemble of stable granular  packings, two levels of randomness are
generally  present  \cite{Brev}.  First,  the force  geometry  clearly
depends  on  the  underlying  geometrical contact  network,  which  is
different in different packings. Secondly, randomness in the values of
the  forces is  present even  in a  fixed contact  network,  since the
forces  are  not  necessarily  uniquely determined  from  the  contact
network.   Instead   of   considering   both  levels   of   randomness
simultaneously, a  natural first step  is thus to obtain  the averages
for a  fixed contact geometry, and  then possibly to  average over the
contact geometries.

\begin{figure}[h]
\begin{center} 
\begin{minipage}{0.4\linewidth}
\includegraphics [width=0.8\linewidth]{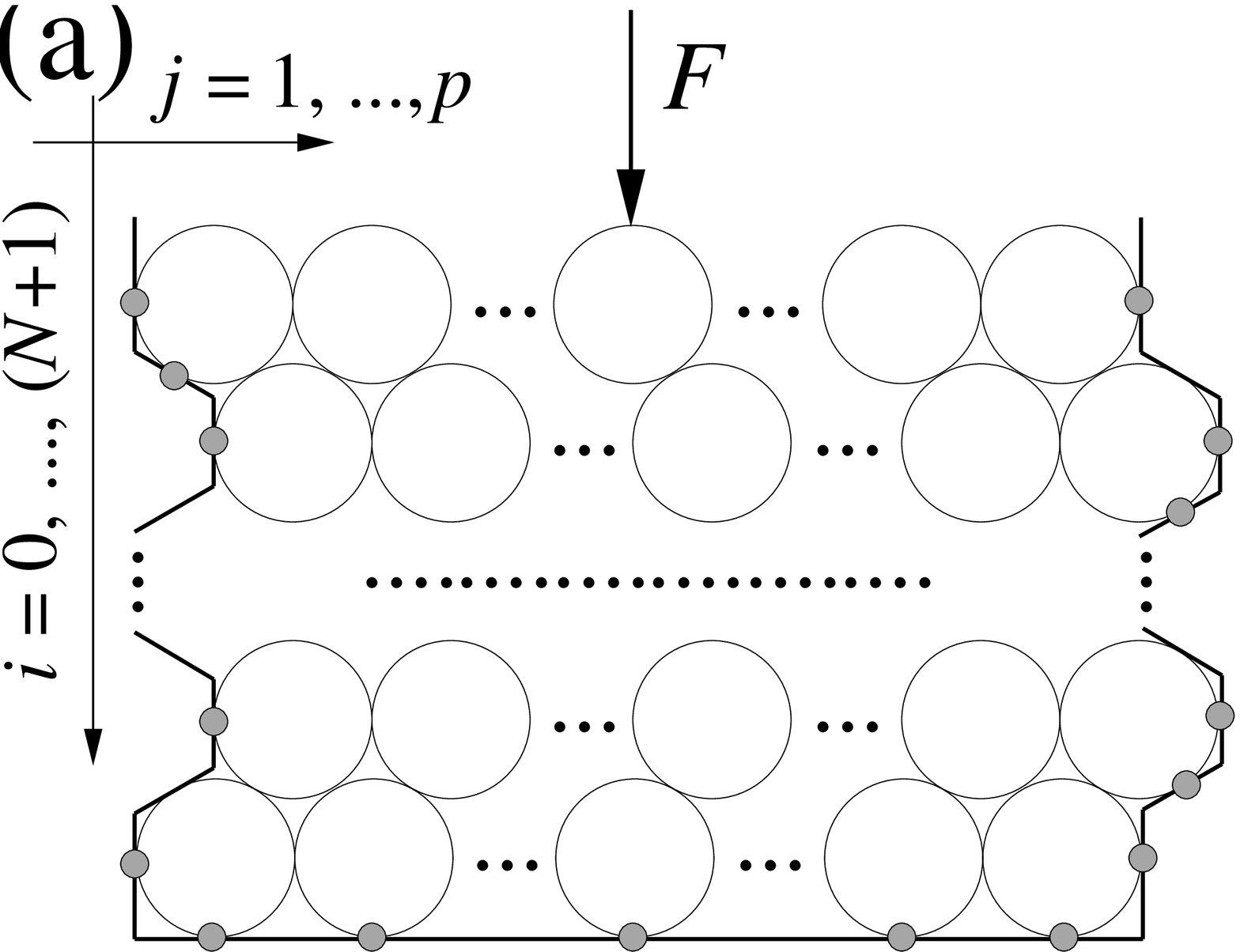}
\end{minipage}
\begin{minipage}{0.58\linewidth}
\includegraphics[width=0.9\linewidth]{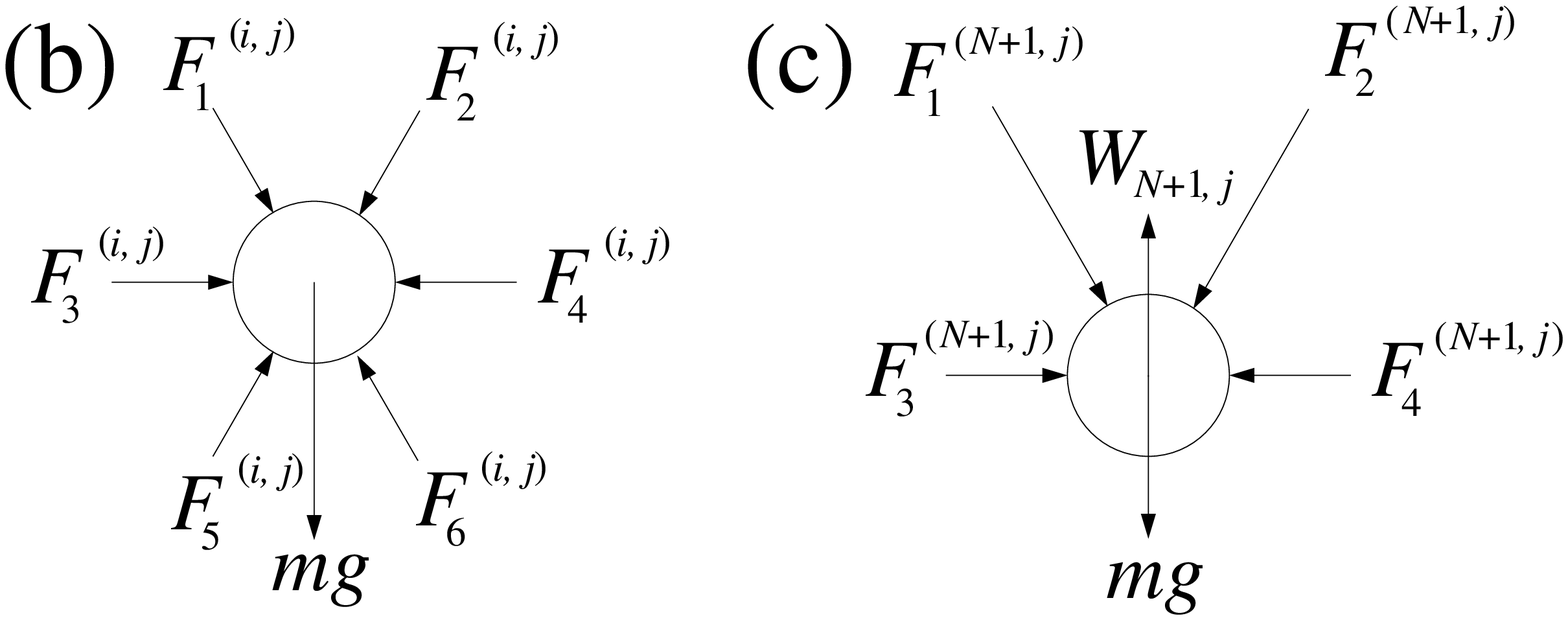}
\end{minipage}
\caption{(a)The model: $(N+2)\times p$ array of hexagonally close-packed
rigid frictionless  spherical grains in two-dimensions  (drawn for odd
$N$). At  the top,  there is only  a single vertically  downward point
force applied on one particle.  At the boundaries, little gray circles
appear on interfaces where the contact forces are non-zero. (b-c) Schematically shown forces on the $j$th grain in the $i$th
layer: (b) $i\leq N$, (c)  $i=(N+1)$, the bottom reaction $W_{N+1}$ is
shifted upwards for clarity; $F^{(i,j)}_m\geq0\,\,\forall m$. }
\label{fig1}
\end{center}
\end{figure}

While  such  a  method  has  recently been  shown  to  produce  single
inter-grain  force probability  distributions in  fixed  geometry that
compare  well  with  experiments   \cite{jacco},  in  this  Letter  we
demonstrate  that  it  also  leads  to an  average  response  function
qualitatively  in  agreement  with  experiments.  More  precisely,  we
determine the behaviour of the response of a two-dimensional hexagonal
packing  of rigid,  frictionless spherical  grains placed  between two
vertical  walls (see  Fig.\ref{fig1}),  due to  a vertically  downward
force $F$ applied  on a single grain at  the top layer. Experimentally
\cite{exp1a, exp1b} it  was found that a force $F$  applied to the top
of  a hexagonal  packing of  photo-elastic particles  propagates mainly
along  the two  downward  lattice directions.  We  define  the
response  of  the packing  as  $\left[\langle W_{i,j}\rangle  -\langle
W_{i,j}^{(0)}\rangle\right]/F$ where $W_{i,j}$ and $W_{i,j}^{(0)}$ are
the vertical  force transmitted  by the $(i,j)$th  grain to  the layer
below it respectively with and without the external force $F$, and the
angular   brackets  denote  averaging   over  all   configurations  of
mechanically stable contact forces with equal probability. 


To  start  with,  we  describe  a method  for  assigning  the  uniform
probability  measure  on  the  ensemble  ${\cal  E}$  of  {\it  stable
repulsive  contact   forces\/}  pertaining  to   a  fixed  geometrical
configuration of  $P$ {\it rigid,  frictionless two-dimensional grains
of  arbitrary   shapes  and  sizes\/}  (for   a  rigorous  geometrical
description  of a  granular  packing, see  Ref. \cite{edwards2}).  The
directions of the forces are fixed  at each of the $Q$ contact points,
and  one can  represent any  force  configuration by  a column  vector
$\mathsf F$ consisting of  $Q$ non-negative scalars $\{F_{k} \}$ (with
${k=1,\ldots, Q}$) as its  individual elements. These elements satisfy
$3P$ Newton's  equations ($3$ equations  per grain: two  for balancing
forces in  the $x$ and $y$  directions and one  for balancing torque),
which  can  be represented  as  ${\mathbf A}\cdot{\mathsf  F}={\mathsf
F}_{{ext}}$.  Here,  $\mathbf{A}$  is   a  $3P\times  Q$  matrix,  and
$\mathsf{F}_{{ext}}$ is a  $3P$-dimensional column vector representing
the external forces.  If we  assume $3P<Q$ \cite{usual}, then there is
no unique solution for $\mathsf{F}$. Instead, there exists a whole set
of solutions  that can be  constructed via the three  following steps:
(1) one first identifies  an orthonormal basis $\{{\mathsf F}^{(l)}\}$
($l=1,\ldots,d_K=Q-3P$)  that spans the  space of  $Ker({\mathbf A})$;
(2) one  then  determines a  unique  solution  ${\mathsf F}^{(0)}$  of
${\mathbf  A}\cdot{\mathsf F}^{(0)}={\mathsf F}_{{ext}}$  by requiring
${\mathsf F}^{(0)}.{\mathsf F}^{(l)}=0$  for $l=1,\ldots,d_K$; and (3)
one  finally  obtains   all  solutions  of  ${\mathbf  A}\cdot{\mathsf
F}={\mathsf       F}_{{ext}}$      as       ${\mathsf      F}={\mathsf
F}^{(0)}+\sum\limits_{l=1}^{Q-3P}f_l\,{\mathsf  F}^{(l)}$, where $f_l$
are real numbers. This implies  that ${\cal E}$ is parametrized by the
$f_l$'s  belonging to  a  set ${\cal  S}$  obeying the  non-negativity
conditions for all forces. The uniform measure on ${\cal E}$, which is
usually  compact  \cite{tamas},  is  thus equivalent  to  the  uniform
measure $d\mu=\prod\limits_{k}  dF_k\, \delta({\mathbf A}\cdot{\mathsf
F}-{\mathsf  F_{{ext}}}   )\,\Theta({F_k})=\prod\limits_{l}  df_l$  on
${\cal S}$.

In  our model,  the grains  are spherical,  so that  the  dimension of
${\mathbf A}$ reduces to $2P\times  Q$.  We consider the force $F$ and
the  weights of  the individual  grains  as the  non-zero elements  of
${\mathsf  F}_{{ext}}$,  while  ${\mathsf   F}$  is  composed  of  all
inter-particle  and  non-zero boundary  forces.  Simple counting  then
shows that $Q=3Np+5p+N+2$ (see  Fig. \ref{fig1}). The matrix ${\mathbf
A}$ represents two equations per particle (see Fig.  \ref{fig1} (b-c))
\cite{usual}
\begin{eqnarray}
F^{(i,j)}_5=F^{(i,j)}_2+mg/\sqrt{3}+[F^{(i,j)}_4-F^{(i,j)}_3]\nonumber\\
F^{(i,j)}_6=F^{(i,j)}_1+mg/\sqrt{3}-[F^{(i,j)}_4-F^{(i,j)}_3]\nonumber\\
W_{N+1,j}=\sqrt{3}\left[F^{(N+1,j)}_1+F^{(N+1,j)}_2\right]/2+mg\nonumber\\
F^{(N+1,j)}_4-F^{(N+1,j)}_3=\left[F^{(N+1,j)}_1-F^{(N+1,j)}_2\right]/2\,,
\label{force_balance}
\end{eqnarray}
i.e.,    $2(N+2)p$    equations    all   together,    implying    that
$d_K=N+2+(N+1)p$.  We  choose $F^{(i,1)}_3$'s  for  $i\leq (N+1)$  and
$F^{(i,j)}_4$'s  for $i\leq  N,1\leq j\leq  N$ to  parameterize ${\cal
E}$.  Once  these  forces  are  fixed, all  the  others  are  uniquely
determined by  solving Eq.  (\ref{force_balance}) layer by  layer from
top  down \cite{equal}. It  is easily  seen that  the number  of these
parameters  is  indeed  $d_K$,  as  it should  be.  Clearly,  in  this
formulation,
$W_{i,j}=\sqrt{3}\left[F^{(i,j)}_1+F^{(i,j)}_2\right]/2+mg$,        and
$d\mu=\prod\limits_{i=0}^{N+1}
dF^{(i,1)}_3\!\!\prod\limits_{(i',j)=(0,1)}^{(N,p)}\!\!dF^{(i',j)}_4$
on     ${\cal    S}$    for     our    model.     Furthermore,    with
$G_{i,j}=F^{(i,j)}_4-F^{(i,j)}_3$, i.e., with \cite{equal}
\begin{eqnarray}
F^{(i,j)}_4=F^{(i,1)}_3+{\textstyle\sum\limits_{j'=1}^j}G_{i,j'}\,,
\label{FGtransform}
\end{eqnarray}
$\prod\limits_{(i,j)=(0,1)}^{(N,p)}\!\!dF^{(i,j)}_4$ in  $d\mu$ can be
replaced by  $\prod\limits_{(i,j)=(0,1)}^{(N,p)}\!\!dG_{i,j}$. In this
form of $d\mu$, in order  to respect the non-negativity conditions for
$F^{(i,j)}_5$'s and $F^{(i,j)}_6$'s, one must satisfy
\begin{eqnarray}
\displaystyle{-\left[F^{(i,j)}_2+mg/\sqrt{3}\right]\leq  G_{i,j}  \leq
F^{(i,j)}_1+mg/\sqrt{3}}\,,
\label{Gineq}
\end{eqnarray}
implying that the set ${\cal  S'}$ of allowed values of $G_{i,j}$'s is
compact.   However,   since    the   non-negativity   conditions   for
$F^{(i,j)}_3$  and  $F^{(i,j)}_4$'s  provide  only  lower  bounds  for
$F^{(i,1)}_3$'s, ${\cal S}$ in this model is actually unbounded.

\begin{figure}

\begin{minipage}{0.5\linewidth}
\begin{center}
\includegraphics[width=0.6\linewidth]{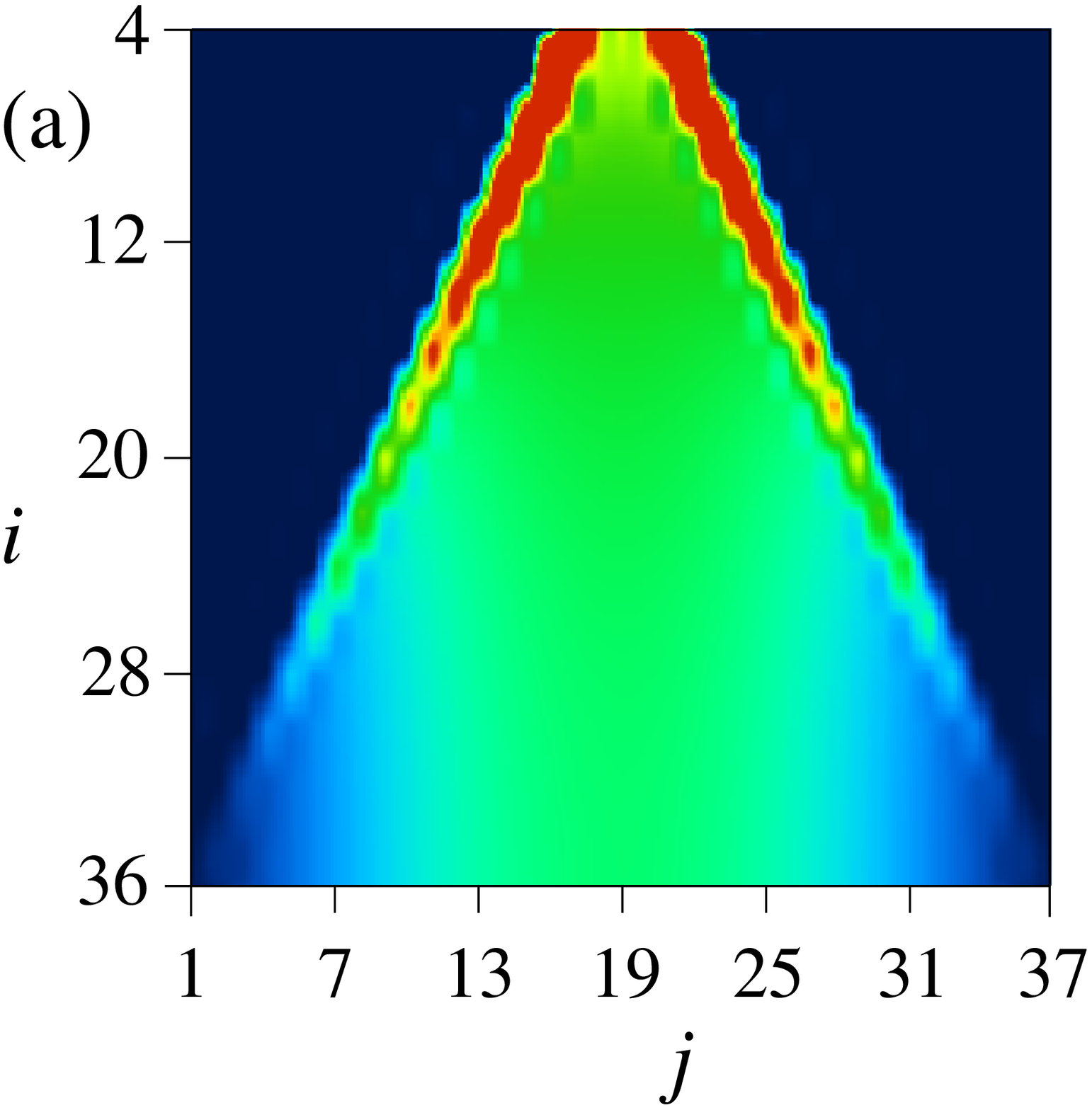}
\end{center}
\end{minipage}
\begin{minipage}{0.5\linewidth}
\begin{center}
\includegraphics[width=0.6\linewidth]{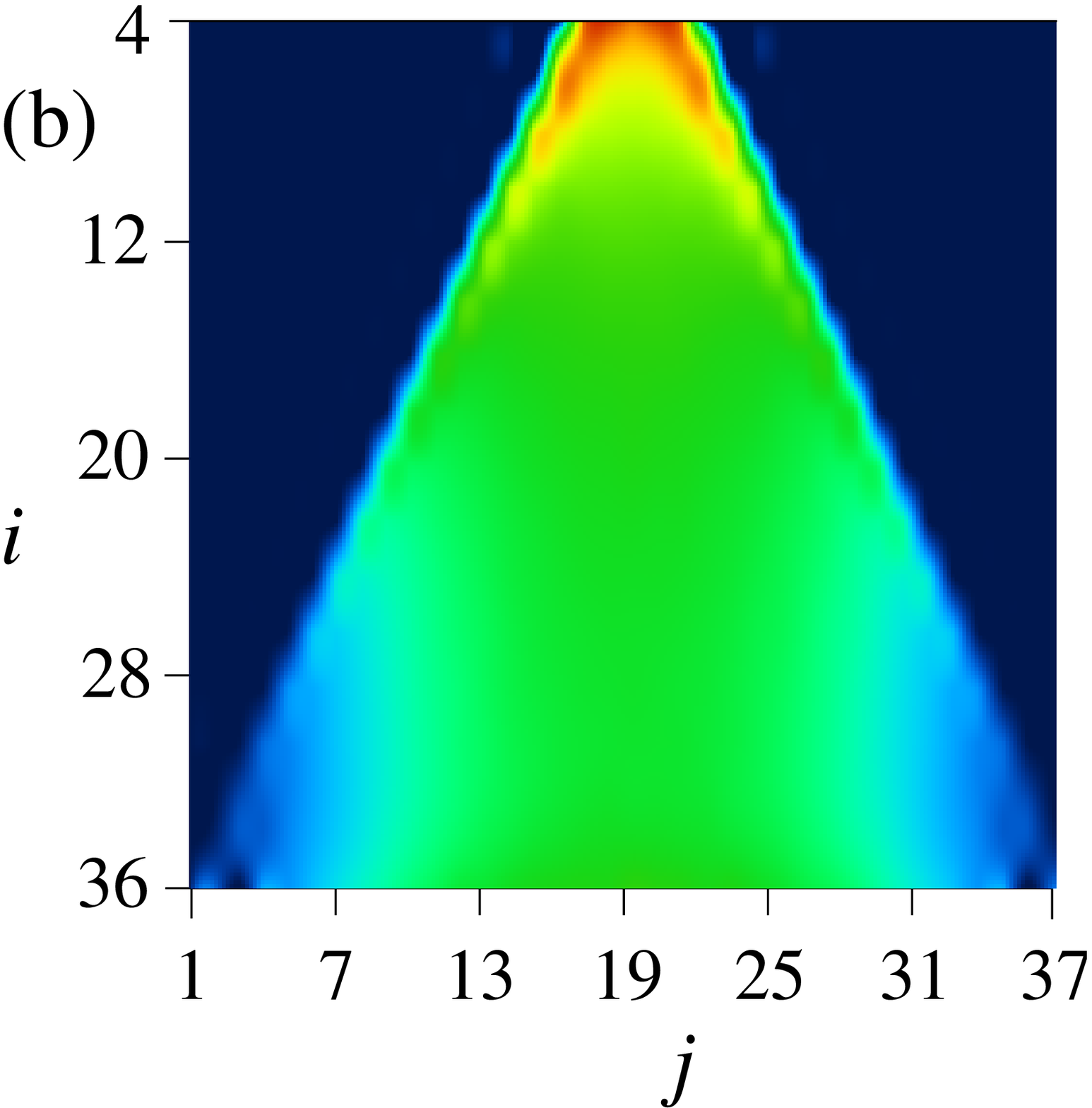}
\end{center}
\end{minipage}
\caption{Colour plots for $N=35$ and $m=0$: (a) mean response; (b) the standard
  deviation of the response. \label{colour}}
\end{figure}

\begin{figure}
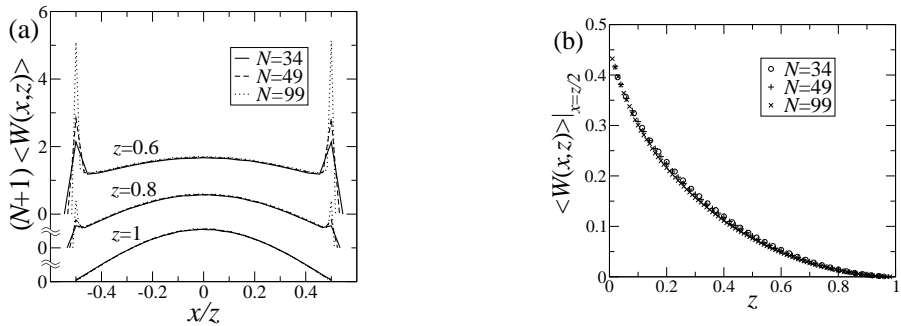

\begin{minipage}{0.5\linewidth}
\begin{center}
\includegraphics[width=0.65\linewidth]{fig4}
\end{center}
\end{minipage}
\begin{minipage}{0.5\linewidth}
\begin{center}
\includegraphics[width=0.65\linewidth]{fig5}
\end{center}
\end{minipage}
\caption{ Behavior of $\langle W(i,j)\rangle$ for $m=0$ in reduced
co-ordinates $x$ and $z$:  (a) scaling of $\langle W(x,z)\rangle$ with
system size for $|x|<z/2$ at  two $z$ values (for clarity, $z=0.8$ and
$z=0.6$ have been shifted upwards  by one and two units respectively);
(b) data  collapse for  $\langle W(x,z)\rangle|_{x=z/2}$ at  three $N$
values. See text for further details.\label{weightless}}
\end{figure}

The remedy we  use is to fix the $F^{(i,1)}_3$  values: indeed, as can
be  seen in  Eq. (\ref{force_balance}),  the values  of  the $W_{i,j}$
depend only the $G_{i,j}$'s so that in this model the precise values of
$F^{(i,1)}_3$ have  no physical meaning.  Nevertheless, one has  to be
careful:  notice that  the $G_{i,j}$'s  are {\it  differences}  of the
physical contact forces  and thus they are allowed  to become negative
in  magnitude.  In fact,  Eqs.  (\ref{FGtransform}) and  (\ref{Gineq})
together imply that if $F^{(i,1)}_3 < 2[F+(i-1)mg]/\sqrt{3}$, then the
positivity requirement of the $F^{(i,j)}_4$ might further restrict the
choice   of  $G_{i,j}$   values  within   the  bounds   of  inequality
(\ref{Gineq}).     In   this   Letter,    we   fix    the   magnitudes
$F^{(i,1)}_3=2[F+(i-1)mg]/\sqrt{3}\equiv F_0$ so that {\it all} values
of $G_{i,j}$ within the bounds of inequality (\ref{Gineq}) are allowed
(details of  the cases $F^{(i,1)}_3<F_0$  appear elsewhere \cite{calc,
pg}) This arrangement  reduces the uniform measure over  ${\cal E}$ to
the uniform  measure on ${\cal  S'}$, which is  a $(N+1)p$-dimensional
polyhedron.

To  evaluate  $\langle  W_{i,j} \rangle=  \displaystyle{\frac{1}{{\cal
N}}\int}_{\!\!\cal  S'}  \,W_{i,j}\prod_{kl}  dG_{k,l}$, where  ${\cal
N}=\int_{{\cal S}'}\prod_{ij}dG_{i,j}$  is the normalization constant,
we define
\begin{eqnarray}
q_{i,j}=\left[\sqrt{3}(G_{i,j}+F^{(i,j)}_2)/2+mg/2\right]/W_{i,j}\,,
\label{varchange}
\end{eqnarray}
where  $q_{i,j}$  is the  fraction  of  $W_{i,j}$  that the  $(i,j)$th
particle transmits to  the layer below it toward  the left, i.e., $F^
{(i,j)}_5=  2q_{i,j}  W_{i,j}/\sqrt{3}$ and  $F^{(i,j)}_6=2(1-q_{i,j})
W_{i,j}/\sqrt{3}$.    Equation    (\ref{varchange})    then    reduces
Eq. (\ref{Gineq}) to  $0 \leq q_{i,j} \leq 1$.  Clearly, $W_{0,j}$ are
the external forces applied on  the top layer. For $i>0$, $W_{i,j}$ is
a function of $q_{k,l}$ for $k<i$, since
\begin{eqnarray}
W_{i,j}=(1-q_{i-1,j-1})\,W_{i-1,j-1}+q_{i-1,j}\,W_{i-1,j} +mg.
\label{qprop}
\end{eqnarray}

It may seem  from Eq. (\ref{qprop}) that in  the hexagonal geometry of
Fig.    \ref{fig1},     one    simply    recovers     the    $q$-model
\cite{qmodel}. There  is however an important subtlety  to take notice
of. In the $q$-model, the  $q$'s corresponding to different grains are
usually uncorrelated, while in our case, the uniform measure on ${\cal
S}'$ implies, from Eq. (\ref{varchange}), that
\begin{equation}
\textstyle{\prod\limits_{i,j}}\,dG_{i,j}=2^{(N+1)p}\,\textstyle{\prod\limits_{i,j}}
dq_{i,j}\,W_{i,j}(q)\big/3^{(N+1)p/2}\,.
\label{jac}
\end{equation}
Due  to  the presence  of  the  Jacobian on  the  right  hand side  of
Eq. (\ref{jac}),  the uniform measure  on ${\cal S'}$ translates  to a
non-uniform measure  on the  $(N+1)p$-dimensional unit cube  formed by
the accessible values of the $q$'s.

Notice an  important artifact of this approach:  the joint probability
distribution $\prod\limits_{i,j} W_{i,j}(q)$  depends on the $q_{i,j}$
values  over  the {\it  whole\/}  system,  thereby making  $q_{i,j}$'s
correlated  with   each  other.  In  fact,   the  induced  probability
$P(q_{i,j})$ for  a {\it single\/} grain  does not only  depend on the
number of  layers present above the  grain, but also on  the number of
layers {\it below\/} it. It is thus clear that the forces in this model
do not  propagate top down,  as they do in  hyperbolic ``stress-only''
models \cite{Brev}.

For  massless  grains  ($m=0$),  it  is  clear  that  (i)  for  $F=0$,
$W_{ij}^{(0)}=0\,\forall(i,j)$   (ii)  the   $\langle  W_{i,j}\rangle$
values  scale linearly  with  $F$  (hence, we  use  $F=1$), and  (iii)
$\langle  W_{i,j}\rangle=0$ outside  the  triangle formed  by the  two
downward lattice directions  emanating from $(i,j)=(0,j_0)$, the point
of  application  of $F$.   The  $\langle W_{i,j}\rangle$'s,  evaluated
numerically  via the Metropolis  algorithm on  these $q$'s,  appear in
Fig. \ref{weightless}.

Our simulation  results for $\langle W_{i,j}\rangle$  and the standard
deviation $\delta W_{i,j}=\sqrt{\langle
W^2_{i,j}\rangle-\langle W_{i,j}\rangle^2}$ within the
triangle are  plotted in Fig. \ref{colour},  using the built-in
cubic  interpolation function  of Mathematica.  Outside  the triangle,
$\langle  W_{i,j}\rangle\equiv0$ appears in  deep indigo;  the largest
$\langle  W_{i,j}\rangle$ value  within the  triangle appears  in dark
red;  and  any  other   non-zero  $\langle  W_{i,j}\rangle$  value  is
represented    by    a   linear    wavelength    scale   in    between
\cite{interpol}.   We  find   $\forall  N$   that  (a)   the  $\langle
W_{i,j}\rangle$  values  display  two {\it  single-grain-diameter-wide
symmetric  peaks\/} that  lie precisely  on the  two  downward lattice
directions  emanating from $j_0$,  (b) the  magnitudes of  these peaks
decrease  with depth,  and (c)  only  a central  maximum for  $\langle
W_{i,j}\rangle$  is  seen at  the  very  bottom  layer ($i=N+1$).  The
standard deviation $\delta W_{ij}$ has a similar shape to $\langle
W_{i,j}\rangle$, although the peaks are less pronounced.

We   further    define   $x=(j-j_0)/(N+1)$   and   $(j-j_0+1/2)/(N+1)$
respectively   for   even   and   odd  $i$,   and   $z=i/(N+1)$   [see
Fig. \ref{fig1}] in  order to put the vertices  of the triangle formed
by  the  locations  of  non-zero $\langle  W_{i,j}\rangle$  values  on
$(0,0),  (-1/2,1)$  and  $(1/2,1)\,\,\forall  N$. The  excellent  data
collapse shown  in Fig.  \ref{weightless} indicates that  the $\langle
W_{i,j}\rangle$  values for  $|x|<z/2$ scale  with the  inverse system
size  [Fig.  \ref{weightless}(a);  we  however  show  only  three  $z$
values], while  the $\langle W_{i,j}\rangle$ values  for $|x|=z/2$ lie
on the  same curve for  all system sizes  [Fig.  \ref{weightless}(b)].
The data suggest that in the thermodynamic limit $N\rightarrow\infty$,
the {\it  response field\/} $\langle W(x,z)\rangle$  scales $\sim 1/N$
for  $|x|<z/2$,  but reaches  a  {\it  non-zero\/}  limiting value  on
$|x|=z\,\,\forall         z<1$.          We        thus         expect
$\lim\limits_{N\rightarrow\infty}\langle
W(x,z)\rangle\big|_{|x|=z/2}>\langle
W(x,z)\rangle\big|_{|x|<z/2}\,\,\forall z<1$;  or equivalently, a {\it
double-peaked response field at all depths $z<1$ for large $N$}. 

We have not  found a simple explanation for  such scaling behaviour of
$\langle  W(x,z)\rangle$.  It however  turns  out  that {\it  exact\/}
analytical   expressions  can   be   obtained  for   all  moments   of
$W_{i,j}\,\,\forall(i,j)$,  for any  $N$.   The detailed  calculations
appear in Ref. \cite{calc}.

\begin{figure}[h]
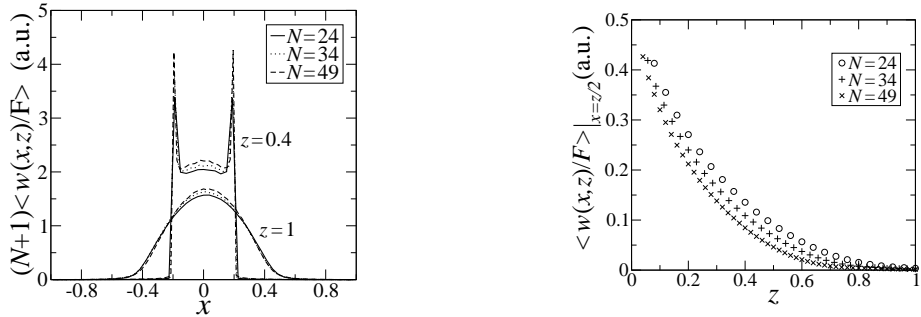
 
\begin{minipage}{0.5\linewidth}
\begin{center}
\includegraphics[width=0.65\linewidth]{fig6}
\end{center}
\end{minipage}
\hspace{2mm}
\begin{minipage}{0.5\linewidth}
\begin{center}
\includegraphics[width=0.65\linewidth]{fig7}
\end{center}
\end{minipage}
\caption{Scaling properties  of $\langle w(x,z)\rangle$,  analogous to
Figs.  \ref{weightless}(b-c),  for $\beta=100$  and  three $N$  values
(a.u.$\,\equiv\,$arbitrary units).}
\label{fig4}
\end{figure}
 In  view of the  self-similarity of  $\langle W(x,z)\rangle$  that we
observe for different system  sizes in Figs. \ref{weightless}(b-c), it
seems natural that we also  study the same properties for $m\neq0$. In
this   case,    $\langle   W_{i,j}^{(0)}\rangle\neq0$   and   $\langle
W_{i,j}\rangle\not\propto F$. To minimize the effect of the boundaries
in the  regions around $j=j_0$,  we have used $p=2N+5$.  For $m\neq0$,
the   relevant  scale   for  the   magnitude  of   $F$   is  obviously
$\alpha=F/mg$.   For $\alpha\neq0$,  {\it  just like  in  the case  of
$m=0$, we  observe a double-peaked  response, and the peaks  are still
single grain diameter wide}. Furthermore, for a given value of $N$ and
increasing $\alpha$,  the magnitude of the response  on $x=z/2$ decays
more  slowly, i.e. the  peaks penetrate  the packing  to progressively
higher values  of $z$.  In order  to avoid repetition,  we do  not use
colour  figures  like  Fig.  \ref{weightless}(a) to  demonstrate  this
behaviour, but the trend of  the data clearly indicates that for fixed
$N$,  one should  recover the  results corresponding  to $m=0$  in the
limit $\alpha\rightarrow\infty$.

 It is  clear from  the qualitative behaviour  described in  the above
paragraph that  in order  to obtain scaling  with increasing  $N$, one
also  needs to  scale $\alpha$  in some  way. To  this end,  we define
$\beta=\alpha/(N+1)$ and keep $\beta$ constant for increasing $N$. The
corresponding graphs are shown in Fig. \ref{fig4} for $\beta=100$. The
fact that the self-similarity  in Fig. \ref{fig4} for different system
sizes is  not as striking  as in Figs.  \ref{weightless}(b-c) suggests
that there  is more to the  story of scaling properties.  It is likely
that the full scaling properties can be unraveled only at much higher
values  of  $N$,  but   unfortunately,  simulations  with  $N$  values
significantly higher than $50$ require impractically long times.

In  summary,   we  find  that  assigning  equal   probability  to  all
mechanically  stable  force  configurations  for  rigid,  frictionless
spherical  grains   (with  or  without  mass)   in  a  two-dimensional
hexagonally close-packed geometry yields a double-peaked response. The
peaks are  single grain  diameter wide, they  lie on the  two downward
lattice  directions  emanating  from   the  point  of  application  of
$F$. With increasing  depth, the magnitude of the  peaks decreases, and a
third peak  starts to develop  directly below the applied  force. Near
(and on)  the bottom  layer only the  middle peak persists;  i.e., the
response becomes single-peaked. As  the number of layers is increased,
the transition  from double to single  peak takes place  deeper in the
packing.  Moreover, for grains each with mass $m$, the peaks penetrate
the  packing deeper  with larger  $F$. The  standard deviation  of the
response is similar in shape to the response, but the peaks are weaker.

We  emphasize that  the results  presented here  are obtained  for the
boundary condition  $F^{(i,1)}_3\geq F_0$. The  case $F^{(i,1)}_3<F_0$
and other  kinds of boundary  conditions have been  analyzed elsewhere
\cite{calc,   pg}.    These  results   together   indicate  that   the
quantitative behaviour  of the response depends crucially  on the side
forces (i.e. boundary conditions)  --- this feature is consistent with
other theoretical approaches \cite{gold}.  In particular, we note that
the transition  to a  single-peaked response does  not take  place for
$F^{(i,1)}_3$ sufficiently small.


We also note that the double-peaked structure of both the mean response and the standard
deviation of the response is in qualitative agreement with experiments
\cite{exp1a, exp1b, exp2}, but the fluctuations observed in the
model are much weaker than found in experiments \cite{exp1b}. 
Another crucial difference between our and the experimental results is
that in this model the peaks are single-diameter wide, while in experiments
the peaks widen with depth \cite{exp1b}.
This  difference probably stems  from the
fact that in experiments the effect of inter-grain friction can never be neglected. 
Presence  of friction would  also certainly  make fluctuations  in the
response $\delta W_{i,j}$ stronger. A study of the effects of friction
on the response along the lines of \cite{clement}
is therefore an important direction for future work.

It  is a  pleasure to  thank J.-P.  Bouchaud, D.  Dhar, J.  M.  J. van
Leeuwen,   B.  Nienhuis,   J.  Snoeijer   and  D.   Wolf   for  useful
discussions.  Financial support  was  provided by  the Dutch  research
organization FOM (Fundamenteel Onderzoek der Materie).


\begin{thebibliography}{03}
\bibitem{usual} For frictionless rigid spherical grains, the number of
forces    is   exactly    equal   to    the   number    of   equations
\cite{mouk}. Numerical  studies \cite{silb}  however find that  in the
limit of large  but finite rigidity the number  of equations is always
less then the number of contacts, and this is what we assume here.
\bibitem{equal} One uses the fact that $F^{(i,j)}_4=F^{(i,j+1)}_3$ for
$1\leq j<p$.
\bibitem{interpol}  The thin green  regions that  appear on  the outer
edge of  the triangle are artifacts  of the interpolation.  It is also
important  to note  that the  horizontal  $\langle F^{(i,j)}_4\rangle$
forces   are  non-zero  everywhere,   but  they   are  not   shown  in
Fig. \ref{weightless}(a).
\end{thebibliography}
\end{document}